\documentclass[aps,twocolumn,showpacs,nofootinbib,amssymb]{revtex4-1}
\date{\today}
\usepackage{graphicx}
\usepackage{color}
\usepackage{amsmath}
\usepackage{amsfonts}
\usepackage{amssymb}

\begin{document}

\title{Partial conservation of seniority in $j=9/2$ shell: Analytic and numerical studies}

\author{Chong Qi}
\thanks{Email: chongq@kth.se}
\affiliation{KTH (Royal Institute of Technology), Alba Nova University Center,
SE-10691 Stockholm, Sweden}

\begin{abstract}
Recent studies show that for systems with four identical fermions in the $j=9/2$ shell two special states, which have seniority $v=4$ and total spins $I=4$ and 6, are eigenstates of any two-body interaction. These states have good seniority for an arbitrary interaction.
In this work an analytic proof is given to this peculiar occurrence
of partial conservation of seniority which is the consequence
of the special property of certain coefficients of fractional parentage.
Further calculations did not reveal its existence in systems with other $n$ and/or $I$ for shells with $j\leq15/2$.
\end{abstract}


\maketitle
\section{Introduction}

It is well known that seniority remains a good quantum number for
systems with identical fermions in a single-$j$ shell when $j\leq7/2$,
irrespective of (rotationally invariant) interactions used. This property of seniority
conservation is no longer valid in shells with $j\geq9/2$. To conserve seniority, the acting two-body interaction has to satisfy $[(2j-3)/6]$
linear constraints ($[n]$ denotes the largest integer not exceeding $n$). For example,  for $j=9/2$ shell the number of conditions is $[(2j-3)/6]=1$ and the necessary and sufficient condition for the conservation of seniority is~\cite{Talmi93}
\begin{equation}
65V_2-315V_4+403V_6-153V_8=0,
\end{equation}
where $V_J=\langle j^2;J|\hat{V}|j^2;J\rangle$ are two-body matrix elements of the interaction $\hat{V}$. $|j^2;J\rangle$ denotes a two-particle state coupled to angular momentum $J$ which runs over even values from 0 to $(2j-1)$. Such conservation conditions are not satisfied by most general two-body interactions for which the eigenstates would be admixtures of states with different seniorities.
However, it was
noted that in $j=9/2$
shell some special eigenstates have good
seniority for an arbitrary interaction~\cite{Escuderos06,Zamick07}. The states are eigenstates of any two-body interaction and exhibit partial dynamic symmetry and the solvability property ~\cite{Isacker08,Levi10}. The partial conservation of seniority in these states may shed light on the existence of seniority isomers in nuclei in the $0g_{9/2}$ shell~\cite{Isacker08}.

More specifically, for four identical fermions in $j=9/2$ shell, there are three states with total angular momentum $I=4$ and $I=6$. These states may be constructed so that one state has seniority $v=2$  (denoted as
$|j^4,v=2,I\rangle$ in the following) and the other two have seniority $v=4$ (denoted as
$|j^4,\alpha_1,v=4,I\rangle$ and $|j^4,\alpha_2,v=4,I\rangle$ where the index $\alpha$ symbolizes an additional quantum number needed when there are more than one states with a given seniority $v$ and total angular momentum $I$). 
The seniority $v=4$ states are not uniquely defined and any linear combination
of them would result in a new sets of $v=4$ states.
However, in Refs.~\cite{Escuderos06,Zamick07} it was found that one special $v=4$, $I=4$ (and $I=6$) state has the interesting property that it has vanishing matrix elements with the remaining two states orthogonal to it even if an interaction that does
not conserve seniority is used. This indicates that the special state conserves seniority and is an eigenstate of any two-body interaction. Zamick and Van Isacker \cite{Zam08}  examined the consequences of this vanishing of non-diagonal matrix elements and showed that it
can be attributed to the special relation of certain one-particle coefficients of factional parentage (cfp) as
\begin{eqnarray}
\nonumber \frac{[j^4(\alpha_1,v=4,I)jI_5|\}j^5,v=3,I_5=j]}{[j^4(\alpha_2,v=4,I)jI_5|\}j^5,v=3,I_5=j]} \\
=\frac{[j^4(\alpha_1,v=4,I)jI_5|\}j^5,v=5,I_5=j]}{[j^4(\alpha_2,v=4,I)jI_5|\}j^5,v=5,I_5=j]},
\end{eqnarray}
where the states $|j^5,v=5,I_5=j\rangle$ and $|j^5,v=3,I_3=j\rangle$ can be uniquely specified. Above relation, which was noted based on the cfp table of Bayman and Lande \cite{Bayman66}, should be valid for any set of $v=4$ and $I=4$ (and $I= 6$) states but an analytic proof of it is still absent. 

The purpose of this paper is to derive an analytic proof to the partial conservation of seniority in $j=9/2$ shell. Calculations will also be carried out to see if such kind of states exist in systems with other $n$ or $j$.  

Firstly in Section II we give a brief description to the problem based on one-particle and two-particle cfp.
Analytic proof of the special property of one-particle cfp [Eq. (2)] is derived in Section III. In Section IV numerical calculations are carried out to explore the possible existence of other partial seniority-conserved solvable state. A summary is given in Section V.

\section{Vanishing of non-diagonal matrix elements}

The problem has been described in Refs. \cite{Escuderos06,Zamick07,Isacker08,Zam08,Levi10,Talmi10} in a variety of ways and will only be briefly discussed here for completeness.
For a system with $n$ identical fermions in a single-$j$ shell (denoted as $j^n$) the matrix elements of the Hamiltonian can be written as linear combinations
of the interaction terms $V_J$ as,
\begin{equation}
H^I_{\alpha\beta}=n(n-1)/2\sum_JM^I_{\alpha\beta}(J)V_J,
\end{equation}
where $I$ is the total spin of the system and $M(J)$ are symmetric matrices.
In particular, the non-diagonal matrix
elements between states involving different seniorities can be written as,
\begin{equation}
H^I_{v_1v_2}=C^I_{v_1v_2}\left[\sum_{J}a_{\lambda J}V_{J}\right],
\end{equation}
where $C^I_{v_1,v_2}$ denotes a coefficient independent of the interaction. By requiring $H^I_{v_1v_2}$=0 we get the conservation conditions of seniority  [e.g., Eq. (1)] which is known in algebraic forms \cite{Talmi93,Rowe10,Isacker08}. $\lambda$ serves as an additional quantum number when more than one conservation conditions are present \cite{Qi10a,Isacker08}.
The number of seniority conditions can be probed by decomposing the two-body matrix elements $V_J$ into quasispin tensors with rank zero and two \cite{Talmi93,Rowe10}. Since the rank zero tensors and the pairing term of rank two tensors do not mix seniority, the number of seniority conservation conditions is related to the
number of linearly independent quasispin rank two tensors.

As mentioned above, there are two $v=4$ states with $I=4$ (and $I=6$) for the $(9/2)^4$ configuration. The non-diagonal matrix elements invovling the special $I=4$ (and $I=6$) state (denoted as $|j^4,a,v=4,I\rangle$ as in Ref. \cite{Zam08}) satisfy
\begin{equation}
H^I_{2a}\equiv H^I_{ab}\equiv0,
\end{equation}
where $|j^4,b,v=4,I\rangle$ denotes the corresponding orthogonal $v=4$ state. Since the state  $|j^4,a,v=4,I\rangle$ is an eigenstate of any interaction, we should also have
\begin{equation}
M^I_{2a}(J)\equiv M^I_{ab}(J)\equiv0,
\end{equation}
which are valid for any angular momentum $J$.

The special $v=4$ state may be written as a combination of an arbitrary set of $v=4$ states as \cite{Zam08}
\begin{equation}
|j^4,a,v=4,I\rangle=\alpha|j^4,\alpha_1,v=4,I\rangle+\beta|j^4,\alpha_2,v=4,I\rangle,
\end{equation}
where the amplitudes are denoted by  $\alpha$ and $\beta$ which can be easily distinguished from the Greeks which symbol different states. It is trivial to construct a special $v=4$ state (through Eq.~(7)) that satisfies $H^I_{2a}=M^I_{2a}(J)=0$ by taking into account the fact that in $j=9/2$ shell there is only one seniority conservation condition and the non-diagonal matrix elements involving different seniorities are in the form of Eq. (4). 
Inserting Eq.~(7) into Eqs.~(5) and (6), immediately we have,
\begin{equation}
\frac{H^I_{2\alpha_1}}{H^I_{{2\alpha_{2}}}} =\frac{M^I_{2\alpha_1}(J)}{M^I_{{2\alpha_{2}}}(J)} = -\frac{\beta}{\alpha},
\end{equation}
and
\begin{equation}
\frac{H^I_{\alpha_1\alpha_1}-H^I_{\alpha_{2}\alpha_{2}}}{H^I_{\alpha_1\alpha_2}}
= \frac{M^I_{\alpha_1\alpha_1}(J)-M^I_{\alpha_{2}\alpha_{2}}(J)}{M^I_{\alpha_1\alpha_2}(J)}
= \left[  \frac{\alpha}{\beta}-\frac{\beta}{\alpha}  \right].
\end{equation}
Above two relations are sufficient in ensuring that the state $|j^4,a,v=4,I\rangle$ is an eigenstate of any Hamiltonian $H$. It is also an common eigenstate of all matrices $M^I(J)$
\begin{equation}
M^I(J)|j^4,a,v=4,I\rangle=E^I_J|j^4,a,v=4,I\rangle,
\end{equation}
where we have $E^I_0=0$, ${\rm rank}[M^I(0)]=1$ and ${\rm Tr}[M^I(0)]=[j^2(I)j^20I|\}j^4,v=2,I]^2$.

For four identical nucleons in a single-$j$ shell, the state can be written as the tensor product of two-particle states as $|j^2(J)j^2(J');I\rangle$ which are not orthonormal~\cite{Liotta81,Qi10,Zhao03b}.
For a given angular momentum $I (I\neq0)$, the seniority $v=2$ state is unique and can be written as (see, e.g., Ref. \cite{Isacker08})
\begin{equation}
|j^4,v=2,I\rangle = \mathcal{N}_{0I}|j^2(0)j^2(J=I);I\rangle,
\end{equation}
where $\mathcal{N}_{0I}$ is the normalization factor.
One of the seniority $v=4$ states can be written as
\begin{eqnarray}
\nonumber|j^4[JJ'],v=4,I\rangle = \mathcal{N}_{JJ'}|j^2(J)j^2(J');I\rangle\\
 - \mathcal{N}_{JJ'}\langle j^2(J)j^2(J')I|j^4,v=2,I\rangle|j^4,v=2,I\rangle,
\end{eqnarray}
where $J$ and $J'$ are the principal parents. The other $v=4$ state can be constructed through the schmidt orthogonalization procedure in a similar way.
The two-particle cfp  for these states can be constructed with the principal-parent scheme \cite{Talmi63, Isacker08} and be expressed in closed forms in terms of $9j$ symbols. 
The special relations of Eqs. (8) and (9) are demonstrated to be true by symbolic calculations with these expressions of two-particle cfp. However, the final expressions are rather complex and cumbersome and will not be given here for simplicity.

The special $v=4$ states can be derived by diagonalizing the Hamiltonian matrix $H$ or matrix $M$. The two-particle cfp for these special $I=4$ and 6 states are given in Table I and II, respectively. The two-particle cfp of the corresponding orthogonal $v=2$ and 4 states are also listed for comparison.
Although these special $v=4$ states can not be constructed through the principal-parent procedure in a simple manner, we found that their overlaps with the normalized $ |j^4[22],v=4,I=4\rangle$ and $ j^4[24],v=4,I=6\rangle$ states are rather large, i.e.,
\begin{equation}
\langle j^4[22],v=4,I=4|j^4,a,v=4,I=4\rangle = 0.998220,
\end{equation}
and
\begin{equation}
\langle j^4[24],v=4,I=6|j^4,a,v=4,I=6\rangle = 0.997704.
\end{equation}

\begin{table}
  \centering
  \caption{Two-particle cfp $[j^2(J)j^2(J')I|\}j^4,\alpha,I]$ for states $|j^4,a,v=4,I=4\rangle$, $|j^4,b,v=4,I=4\rangle$ and $|j^4,v=2,I=4\rangle$.}\label{ngp}
\begin{ruledtabular}
  \begin{tabular}{ccrrr}
$J$&$J'$&$a$&$b$&$v=2$\\
\hline
  0&  4&        0&        0&        0.316228\\
  2&  2&        0.426954&       -0.025505&       -0.225978\\
  2&  4&        0.254224&       -0.198597&        0.103504\\
  2&  6&       -0.310667&       -0.197568&       -0.225866\\
  4&  4&       -0.239508&       -0.331279&        0.083861\\
  4&  6&        0.141827&        0.224545&       -0.194095\\
  4&  8&       -0.156709&        0.387355&       -0.135919\\
  6&  6&       -0.163753&        0.564526&        0.344947\\
  6&  8&       -0.031625&        0.024706&       -0.437779\\
  8&  8&       -0.565594&       -0.108706&        0.062138\\
  \end{tabular}
  \end{ruledtabular}
\end{table}

\begin{table}
  \centering
  \caption{Same as Table I but for those of the $I=6$ states.}
\begin{ruledtabular}
  \begin{tabular}{ccrrr}
$J$&$J'$&$a$&$b$&$v=2$\\
\hline
  0&  6&        0&        0&       -0.316228\\
  2&  4&       -0.165170&       -0.011211&        0.187931\\
  2&  6&       -0.344596&        0.169404&        0.058551\\
  2&  8&        0.376631&        0.150078&        0.135988\\
  4&  4&       -0.266132&       -0.356461&        0.161496\\
  4&  6&        0.037853&       -0.214021&       -0.287013\\
  4&  8&        0.233036&       -0.380268&        0.364254\\
  6&  6&        0.028697&        0.520092&        0.140619\\
  6&  8&        0.213691&        0.241247&        0.126376\\
  8&  8&        0.387024&       -0.050282&       -0.421439\\
  \end{tabular}
  \end{ruledtabular}
\end{table}

\subsection{The matrices $M^I(J=I)$ in terms of one-particle cfp}

The algebraic expressions of the matrix elements of $H$ and $M$ are rather complex in terms of two-particle cfp or $9j$ symbols. On the other hand, Ref. \cite{Zam08} found that the non-diagonal matrix elements of $M^I(J=I)$ can acquire a simple form in terms of one-particle cfp,
\begin{eqnarray}
\nonumber M^I_{2a}(J=I) =5[j^4(v=2,J)jI_5|\}j^5,v=3,I_5=j]\\
\times [j^4(a,v=4,J)jI_5|\}j^5,v=3,I_5=j],
\end{eqnarray}
and
\begin{eqnarray}
\nonumber M^I_{ab}(J=I) &=&5\sum_{v_5=3,5}[j^4(a,v=4,I)jI_5|\}j^5,v_5,I_5=j]\\
&\times&[j^4(b,v=4,I)jI_5|\}j^5,v_5,I_5=j].
\end{eqnarray}

By requiring $M_{2a}(J)=M_{ab}(J)=0$, we have
\begin{eqnarray}
\nonumber &&[j^4,a,v=4,I)jI_5|\}j^5,v=3,I_5=j]\\
\nonumber &=&[j^4,a,v=4,I)jI_5|\}j^5,v=5,I_5=j]\\
&=&0
\end{eqnarray}
The special relation of Eq. (2) was derived by inserting Eq. (7) to above equation.

The one-particle cfp for these special $I=4$ and 6 states are given in Table III and IV, respectively. The one-particle cfp of the corresponding orthogonal $v=2$ and 4 states are also listed for comparison. 

\begin{table}
  \centering
  \caption{One-particle cfp $[j^3(v_3I_3)jI|\}j^4,\alpha,I]$ for states $|j^4,a,v=4,I=4\rangle$, $|j^4,b,v=4,I=4\rangle$ and $|j^4,v=2,I=4\rangle$.}
\begin{ruledtabular}
  \begin{tabular}{ccrrr}
$v_3$&$I_3$&$a$&$b$&$v=2$\\
\hline
  3&  3/2&       -0.122187&        0.473243&        0.284268\\
  3&  5/2&        0.054772&       -0.388546&        0.181186\\
  3&  7/2&       -0.617040&       -0.064668&        0.176295\\
  3&  9/2&        0&        0.349269&       -0.344932\\
  1&  9/2&        0&        0&        0.612372\\
  3&  11/2&        0.404329&        0.328164&        0.363442\\
  3&  13/2&       -0.614814&        0.203062&       -0.156447\\
  3&  15/2&        0.159749&        0.521118&       -0.243006\\
  3&  17/2&        0.185293&       -0.280021&       -0.381691\\
  \end{tabular}
  \end{ruledtabular}
\end{table}

\begin{table}
  \centering
  \caption{Same as Table III but for those of the $I=6$ states.}
\begin{ruledtabular}
  \begin{tabular}{ccrrr}
$v_3$&$I_3$&$a$&$b$&$v=2$\\
\hline
  3&  3/2&        0.106083&       -0.397464&        0.144841\\
  3&  5/2&       -0.309096&       -0.315750&       -0.246183\\
  3&  7/2&       -0.622541&        0.010517&       -0.017630\\
  3&  9/2&        0&        0.330407&       -0.305511\\
  1&  9/2&        0&        0&       -0.612372\\
  3&  11/2&       -0.205106&        0.313264&       -0.161577\\
  3&  13/2&       -0.408432&        0.102112&        0.263339\\
  3&  15/2&        0.109009&       -0.567540&       -0.442498\\
  3&  17/2&        0.388116&        0.430312&       -0.244520\\
  3&  21/2&        0.366397&       -0.131203&        0.314194\\
  \end{tabular}
  \end{ruledtabular}
\end{table}

As noted in Ref.~\cite{Escuderos06}, the one-particle cfp of states $|j^4,a,v=4,I\rangle$ also exhibit the special property of
\begin{equation}
[j^3(v=3,I_3=j)jI|\}j^4,a,v=4,I]=0,
\end{equation}
which is relevant to Eq.~(17) through the recursion relation of the cfp. From Eq.~(19.31) of Talmi's book \cite{Talmi93}, the following relation holds~\cite{Zam08}
\begin{eqnarray}
\nonumber \frac{[j^4(\alpha_1,v=4,I)jI_5|\}j^5,v=5,I_5=j]}{[j^4(\alpha_2,v=4,I)jI_5|\}j^5,v=5,I_5=j]}\\
=
\frac{[j^3(v=3,I_3=j)jI|\}j^4,\alpha_1,v=4,I]}{[j^3(v=3,I_3=j)jI|\}j^4,\alpha_2, v=4, I]},
\end{eqnarray}
which is equivalent to the relation defined by Eq. (2).

\section{Relations between one-particle cfp}

The seniority scheme can be obtained by introducing states associated with the irreducible representations of group chain $U(2j+1)\supset Sp(2j+1)\supset O(3)$ where $U$, $Sp$ and $O$ denote the unitary, symplectic and orthogonal groups, respectively \cite{Talmi93,Rowe10}. 
In Refs.~\cite{Zamick07,Escuderos06,Zam08}, the one-particle cfp are calculated by using the Bayman-Lande procedure \cite{Bayman66} in which the seniority-classified cfp are obtained iteratively by diagonalizing the $Sp(2j+1)$ and $SU(2j+1)$ Casimir operators.

On the other hand, the one-particle cfp can be factorized into the product of the isoscalar factors of the group chains $U(2j+1)\supset Sp(2j+1)$ and  $Sp(2j+1)\supset O(3)$ \cite{Sun89} (see also Ref. \cite{Wang95}). The isoscalar factor is introduced based on the factorization property of the Clebsch-Gordan Coefficient (Racah's factorization lemma), that is the Clebsch-Gordan Coefficients of a group $S_n$ can be factorized into a Clebsch-Gordan coefficient of the subgroup $S_{n-1}$ and an isoscalar factor specified for the group chain $S_n\supset S_{n-1}$. The isoscalar factors of  group chain $U(2j+1)\supset Sp(2j+1)$ are known as analytic expressions while those of $Sp(2j+1)\supset O(3)$ can be calculated iteratively by a recurrence formula. Correspondingly, the $v\rightarrow v-1$ cfp can be factorized as \cite{Sun89}
\begin{eqnarray}
\nonumber [j^{n-1}(\alpha_1,v-1,J_1)jJ|\}j^{n}\alpha v J]\\
=\sqrt{\frac{v(2j+3-n-v)}{n(2j+3-2v)}}
 R(j,v-1,\alpha_1J_1;nv\alpha J),
\end{eqnarray}
where $R$ is the $Sp(2j+1)\supset O(3)$ isoscalar factor.
With the principal-parent procedure, a state $\alpha$ with total angular momentum $J$ can be written as
\begin{equation}
|j^{n}\alpha v J\rangle =
\sum_{\alpha_1'J_1'}c_{\alpha_1'J_1'}|j^{n}\alpha v(\alpha_1'J_1')
J\rangle,
\end{equation}
and
\begin{eqnarray}
 R(j,v-1,\alpha_1J_1;nv\alpha J)& =&
\sum_{\alpha_1'J_1'}c_{\alpha_1'J_1'}\\
\nonumber &&\times R(j,v-1,\alpha_1J_1;nv\alpha(\alpha_1'J_1') J),
\end{eqnarray}
where $\alpha_1'J_1'$ denote the principal parents. The coefficients $c_{\alpha_1'J_1}$ can be determined by the standard orthnormalization procedure. 

The isoscalar factor $R$ is calculated by  a recurrence formula,
\begin{equation}
R(j,v-1,\alpha_1J_1;nv\alpha(\alpha_1'J_1') J)
=\frac{P(\alpha_1'J_1'\alpha_1J_1J)}{\sqrt{v P(\alpha_1'
J_1'\alpha_1'J_1'J)}},
\end{equation}
where
\begin{eqnarray}
\nonumber P(\alpha_1'J_1'\alpha_1J_1J) =
\delta_{\alpha_1'\alpha_1}\delta_{J_1'J_1} \\
+
\nonumber (-1)^{J+J_1'}(v-1)\sqrt{(2J_1'+1)(2J_1+1)}\\
\times\sum_{\alpha_2
J_2}\left[\left\{
            \begin{array}{ccc}
              j & J_2 & J_1' \\
              j & J & J_1 \\
            \end{array}
          \right\}
          +\frac{(-1)^{v}2\delta_{J_2J}}{(2J+1)(2j+5-2v)}\right]\nonumber\\
\nonumber \times R(j,v-2,\alpha_2J_2;nv-1,\alpha_1 J_1)\\
\times R(j,v-2,\alpha_2J_2;nv-1\alpha_1' J_1').
\end{eqnarray}

To prove the special relation of Eqs. (2) \& (23), we start from  the unique state $|j^5, v=5, J=j\rangle$. It can be easily constructed as the tensor product of any $n=4$ state and the single particle. We may take $J_1'=0$ and $|j^{4},v=4, J=J_1'\rangle$ as the principal parent. For this state
we have,
\begin{eqnarray}
\nonumber && [j^{3}(v=3,J_1)jJ|\}j^{4}, v=4, J=0]\\
\nonumber &=&R(j,v=3,J_1;n=4,v=4,
J=0)\\
&=&\delta_{J_1,j}.
\end{eqnarray}
Taking $|\alpha_1'J_1'\rangle=|j^{4},v=4, J_1'=0\rangle$ and $J=j$ and inserting above relation to Eq. (24), we have
\begin{eqnarray}
\nonumber P(\alpha_1'J_1'\alpha_1J_1J) &=& 4\sqrt{(2J_1+1)}\left[\left\{
            \begin{array}{ccc}
              j & j & 0 \\
              j & j & J_1 \\
            \end{array}
          \right\}\right.\\
         &&\left. -\frac{2}{(2j+1)(2j-5)}\right]\nonumber\\
          &&\times R(j,v=3,j;n=4, v=4,\alpha_1 J_1).
\end{eqnarray}

Immediately the special relation of Eq. (19) can be obtained as
\begin{eqnarray}
\nonumber &&\frac{[j^{4}(v=4,\alpha_1,I)jJ|\}j^{5}, v=5,
J=j]}{[j^{4}(v=4,\alpha_2,I)jJ|\}j^{5},v=5, J=j]}\\
\nonumber &=&\frac{R(j,v=4,\alpha_1,I;n=5, v=5,J=j)}{R(j,v=4,\alpha_2,I;n=5, v=5,J=j)}\\
\nonumber &=&\frac{R(j,v=4,\alpha_1,I;n=5, v=5(J_1'=0)J=j)}{R(j,v=4,\alpha_2,I;n=5, v=5(J_1'=0)J=j)}\\
\nonumber& =&\frac{P(\alpha_1'J_1'\alpha_1I,J=j)}{P(\alpha_1'J_1'\alpha_2I,J=j)}\\
\nonumber& =&\frac{R(j,v=3,j;n=4, v=4,\alpha_1 I)}{R(j,v=3,j;n=4, v=4,\alpha_2 I)}\\
&=&\frac{[j^{3}(v=3,J_1=j)jI|\}j^{4},v=4,\alpha_1I]}
{[j^{3}(v=3,J_1=j)jI|\}j^{4},v=4,\alpha_2I]},
\end{eqnarray}
which is equivalent to Eq. (2) as mentioned before.\footnote{This relation may also be derived by using Rq. (28.24) of Ref.~\cite{Talmi63} with the same principle-parent procedure. Detailed derivations will be presented elsewhere. I thank the referee for pointing this out to me.}

\section{Numerical calculations for other systems}

\begin{table}
  \centering
  \caption{States in $j=9/2$ and 11/2 shells that can not be uniquely defined by the seniority $v$ and total spin $I$.}
\begin{ruledtabular}
  \begin{tabular}{cccc}
Configuration &$ I$ & Dimension & $v$\\
\hline
  $(9/2)^4$&  4, 6  &3&      2, 4\\
  $(11/2)^3$&  9/2, 15/2&2&        3\\
  $(11/2)^4$& 2, 10&3&      2, 4\\
  $(11/2)^4$& 4, 6, 8&4&      2, 4\\
  $(11/2)^4$& 5, 7, 9, 12&2&      4\\
  $(11/2)^5$& 5/2,21/2, 23/2&3&      3, 5\\
  $(11/2)^5$& 7/2, 9/2, 13/2, 17/2, 19/2&4&      3, 5\\
  $(11/2)^5$& 11/2&5&     1, 3, 5\\
$(11/2)^5$& 15/2&5&      3, 5\\
$(11/2)^5$& 25/2&2&      5\\
$(11/2)^6$& 2&4&      2, 4, 6\\
$(11/2)^6$& 7, 9, 12&4&      4, 6\\
$(11/2)^6$& 3, 5&3&      4, 6\\
$(11/2)^6$& 4, 8&6&   2, 4, 6\\
$(11/2)^6$& 6&7&   2, 4, 6\\
$(11/2)^6$& 10&5&   2, 4, 6\\
  \end{tabular}
  \end{ruledtabular}
\end{table}

\begin{table}
  \centering
  \caption{Same as Table V but for shells with $j=13/2$ and 15/2. Only states with particle numbers $n=3$ and 4 are shown for simplicity.}
\begin{ruledtabular}
  \begin{tabular}{cccc}
Configuration &$ I$ & Dimension & $v$\\
\hline
  $(13/2)^3$&  9/2, 11/2, 15/2, 17/2, 21/2&2&        3\\
  $(13/2)^4$&  2, 12&4&        2,4\\
  $(13/2)^4$&  4, 6, 10&5&        2,4\\
  $(13/2)^4$&  5, 7, 9, 11, 14 &3&        4\\
  $(13/2)^4$&  8&6&        2,4\\
  $(13/2)^4$&  12, 16&2&        4\\
  $(15/2)^3$&  9/2-13/2, 17/2-23/2, 27/2&2&        3\\
  $(15/2)^3$&  15/2&3&      1, 3\\
  $(15/2)^4$&  0&3&      0, 4\\
  $(15/2)^4$&  2&4&      2, 4\\
  $(15/2)^4$&  3, 17, 20&2&       4\\
  $(15/2)^4$&  4&6&      2, 4\\
  $(15/2)^4$&  5, 15, 18&3&      4\\
  $(15/2)^4$&  6, 8, 10, 12&7&      2, 4\\
  $(15/2)^4$&  7, 11, 13, 16 &4&      4\\
  $(15/2)^4$&  9&5&      4\\
  $(15/2)^4$&  14&5&     2, 4\\
  \end{tabular}
  \end{ruledtabular}
\end{table}

As mentioned before, the two special $v=4$ and $I=4$ and 6 states have solvable eigenvalues and are eigenstates of any two-body interaction. The eigenvalue can be expanded in terms of two-body matrix elements as (see, e.g., Ref. \cite{Qi10})
\begin{equation}
E_I=C^I_JV_J,
\end{equation}
where $C^I_J=X^TM^I(J)X$ and $X$ are the expansion amplitudes of the wave function. $C^I_J$ are independent of interaction if the state is an eigenstate of any interaction.
Moreover, it can be easily recognized that any eigenstate of any two-body interaction should have a definite seniority since admixtures of states with different seniorities cannot yield an eigenstate of any two-body interaction~\cite{Talmi10}.
It may be interesting to see if such kind of state exist in other systems. Such kind of calculations have been done for systems with $n=4$ in Ref.~\cite{Isacker08} in which no other case was found for $j>9/2$.

To explore the properties of states with other $n$, $I$ and/or $j$, we start from the Hamiltonian matrix of Eq.~(3) with an arbitrary interaction. The Hamiltonian matrix is diagonalized numerically by employing the usual shell model diagonalization procedure~\cite{Qi08} for symplicity. If an state is an eigenstate of any interaction, the energy expression (Eq.~(28)) will be uniquely defined and remain the same with the variation of the two-body matrix elements $V_J$. It provides a simple criterion for the searching of  state that exhibits partial seniority conservation.

There are a few trivial cases exist which are not interested here. For example, 
the state is an eigenstate of any two-body interaction if there  is only one state for a given angular momentum $I$. These trivial cases have been discussed in Refs. \cite{Talmi10,Isacker08} and will not be detailed here for simplicity.

We concentrate on systems that have at least two states for a given total spin $I$, especially states that can not be uniquely defined by the total spin $I$ and seniority $v$. These states are listed in Tables V and VI. Calculations are done for systems with nucleon numbers up to $n=(2j+1)/2$ and $j$ values up to 15/2. Systems with a higher $j$ values, which are of less physical relevance comparatively, can be studied in the same manner.

Calculations did not find any other fermionic system that exhibits partial seniority conservation for shells with $j\leq15/2$.

\section{Summary}

Seniority is not conserved by most general two-body interactions in single-$j$ shells with $j\geq9/2$. However,  recent studies \cite{Escuderos06,Zamick07,Isacker08} show that for systems with four identical fermions in the $j=9/2$ shell two special states, with seniority $v=4$ and total spins $I=4$ and 6, have good seniority for an arbitrary interaction and are eigenstates of any two-body interaction.
This peculiar occurrence
of partial conservation of seniority is the consequence
of special property of certain one-particle cfp of Eqs. (2) \& (19) \cite{Zam08}. 

In this paper, the partial conservation of seniority in $j=9/2$ shell is studied with the help of two-particle cfp and one-particle cfp.
Although the two special $I=4$ and 6 state can not be constructed as a simple form within the principal-parent procedure, it is found that their overlaps with the normalized  $|j^4[22],v=4,I=4\rangle$ and $ |j^4[24],v=4,I=6\rangle$ states are more than 99.5\%. The special relation of Eq. (19) is also analytically proven with the principal-parent procedure.

If any state is the eigenstate of any interaction, the corresponding expression (Eq.~(28)) will be uniquely specified and be independent of two-body interactions. Except the special $I=4$ and 6 states mentioned above, calculations did not find its existence in any other fermionic system for shells with $j\leq15/2$.

\section*{Acknowlegement}

This work has been supported by the Swedish Research Council (VR). The author thanks Professor R.J. Liotta for useful discussions.
Partial of this work was done during the author's visit to Professor F.R. Xu of Peking University and Professor Y.M. Zhao of Shanghai Jiaotong University.

\end{document}